\documentstyle[preprint,aps,epsfig]{revtex}
\draft
\begin{document}

\title{Persistent spin currents induced by a spatially-dependent magnetic
field in a spin-1/2 Heisenberg antiferromagnetic ring}

\author{D. Schmeltzer}
\address{Department of Physics, City College of the City University
of New York, New York, NY 10031}
\author{A. Saxena, A. R. Bishop and D. L. Smith}
\address{Theoretical Division, Los Alamos National Laboratory, Los 
Alamos, NM 87545}

\date{\today}

\maketitle

\begin{abstract}
We show that a spatially-dependent magnetic field can induce a
persistent spin current in a spin-1/2 Heisenberg antiferromagnetic
ring, proportional to the solid angle subtended by the magnetic
field on a unit sphere.  The result is a direct consequence of
Berry ``parallel transport" in space.  The magnitude of the spin
current is determined by the ratio of longitudinal and transverse 
exchange interactions $J_\parallel/J_\perp$ and by the magnetic
field.  For large magnetic fields the Zeeman energy strongly
renormalizes the Ising term giving rise to a maximum spin current.
In the limit of $J_\parallel/J_\perp\ll1$ the amplitude of the
current behaves like $1-(J_\parallel/J_\perp)^2$.  In the opposite
limit $\pi J_\perp> J_\parallel>J_\perp$ the amplitude scales as  
$\sqrt{J_\parallel/ J_\perp}\exp{(-\sqrt{J_\parallel/J_\perp})}$.

\end{abstract}

\vskip 0.5truecm 

The generation of persistent currents in a ring is a consequence
of the sensitivity of a wavefunction to its boundary conditions.
The basic idea of persistent currents was put forward a few
decades ago \cite{buttiker,cheung,levy}.  The effect of a magnetic
flux threading a metallic ring, such that the magnetic field at
the ring vanishes, can be replaced by a twist of the
wavefunction boundary conditions.  The change in these boundary
conditions gives rise to an asymmetry between the number of
electrons with $K>K_F$ and $K<-K_F$, where $K_F$ denotes Fermi
momentum.  As a result, an electric current is induced in
mesoscopic ring structures.  For large ring structures, in the
thermodynamic limit, the current vanishes; in agreement with the
fact that no spontaneous broken chirality occurs in
non-interacting systems.  A new interest in this problem has
arisen for non-Fermi liquids. In particular, for Luttinger liquids
novel behavior has recently been proposed \cite{loss,ds1,ds2,ds3}.

In the last ten years the ``parallel transport" idea introduced by
Berry \cite{berry} for adiabatic processes has provided new insight
into currents induced by topological effects.  A typical situation
considered by Berry was a spin-1/2 electron in a time periodic
magnetic field.  Assuming adiabatic motion Berry showed that when
an electron returns to its initial position the wavefunction
accumulates a nontrivial phase equal to the angle of rotation of
the spin-1/2 angular momentum on a sphere.  This is the Berry
phase \cite{berry,simon,nakahara}.

Recently it has been shown \cite{schutz} that for an integer spin
$S=1$ ferromagnet the Berry phase combined with the spin-orbit
effect can give rise to a spin current and a corresponding
electric field.  Here we propose a new way of generating a spin
current, when a spatially periodic magnetic field acts on a {\it
spin-1/2 antiferromagnetic ring a Berry phase in space is induced}
giving rise to a twist on the wavefunction boundary condition.  As
a consequence a persistent spin current is induced.  This result
can be understood within the Berry parallel transport idea since a
spin-1/2 antiferromagnet is Lorentz invariant.

We consider a spin-1/2 antiferromagnetic Heisenberg ring of
circumference $L$ in the presence of a spatially-dependent
magnetic field of the ``crown shape" form, $\vec{h}(x)
=h_0(\sin\theta(x)\cos \phi(x)$, $\sin\theta(x)\sin\phi(x),
\cos\theta(x))$, where $\theta(x)$ and $\phi(x)$ are polar
$-\pi\le\phi(x)\le\pi$, and azimuthal $0<\theta(x)\le\theta_0 
<\pi/2$ angles such that $\phi(x)=\phi(x+L)$, see Fig. 1.  
We restrict $\theta(x)$ to the ``northern"
hemisphere which obeys $\theta(x)=\theta(x+L)$.  The Hamiltonian
for the spin-1/2 antiferromagnet in an external magnetic field is:
$$H=\frac{\tilde{J}_\perp}{2}\sum_x(S_+(x)S_-(x+1)+S_-(x+1)S_+(x))
+\tilde{J}_\parallel\sum_x S_3(x)S_3(x+1)+g\mu_B\sum_x\vec{h}(x)
\cdot\vec{S}(x), \eqno(1) $$ where $\tilde{J}_\perp$ and
$\tilde{J}_\parallel$ are the transverse and longitudinal exchange
coupling constants.

We rotate $\vec{S}(x)$ in the direction of the magnetic field
$\vec{h}(x)$ using the following unitary transformation:
$$\hat{U}=\Pi_{x=0}^{N-1}\exp\left(-i\frac{\phi(x)}{2}\sigma_3\right)
\exp\left(-i\frac{\theta(x)}{2}\sigma_2\right) , \eqno(2) $$ where
the circumference of the ring is $L=Na$ and $\sigma_2$ and
$\sigma_3$ are Pauli spin-1/2 matrices.  As a result of the
unitary transformation Eq. (2), the Hamiltonian $H$ is transformed
to $\tilde{H}=\hat{U}^\dag H \hat{U}$.  Due to the periodicity of
the angles $\theta(x)$ and $\phi(x)$ the transformed wavefunction
obeys the same boundary condition as the original wavefunction.

The transformed Hamiltonian $\tilde{H}$ takes the form,
$$\tilde{H}\simeq\frac{J_\perp}{2}\sum_{x=0}^{N-1}\left(S_+(x)
e^{iA(x,x+1)}S_-(x+1)+S_-(x)e^{-iA(x,x+1)}S_+(x+1)\right) $$
$$+J_\parallel\sum_{x=0}^{N-1}S_3(x)S_3(x+1)
+g\mu_Bh_0\sum_x S_3(x) . \eqno(3) $$ In Eq. (3) $A(x,x+1)$ is the
Berry connection:
$$A(x,x+1)\equiv A_\phi d\phi=\frac{1}{2}(1+\cos\theta(x))\frac{d\phi}
{dx}dx .  \eqno(4) $$ To obtain Eq. (4) we have restricted
$\theta(x)$ to the range $0<\theta(x)<\theta_0\le\pi/4$, where
$\theta_0$ denotes the maximum tilt angle of the ``crown-shaped"
magnetic field (Fig. 1).  The Berry connection results from rotating 
the coordinate of the spin-1/2 operator.  Moving around the ring, a
phase proportional to the subtended solid angle accumulates.

Due to the unitary transformation the exchange couplings
$\tilde{J}_\perp$, $\tilde{J}_\parallel$ are replaced by $J_\perp$
and $J_\parallel$, where $J_\parallel\equiv\tilde{J}_\parallel
\langle\cos^2\theta(x)\rangle$, $J_\perp\equiv\tilde{J}_\perp
((1+\langle\cos^2\theta(x)\rangle)/2)$ with
$\langle\cos^2\theta(x)
\rangle\equiv(1/\theta_0)\int_0^{\theta_0}\cos^2\theta
d\theta=(1/2) (1+(\sin2\theta_0/2\theta_0))$.

Following Ref. [12] we use a Jordan-Wigner transformation to
replace the spin-1/2 $\vec{S}(x)$ operator by the spinless
fermions $c^\dag$, $c(x)$:
$$S_-(x)=e^{-i\pi\sum_{x=0}^{N-1}n(x)}c(x), ~~~S_+(x)=[S_-(x)]^\dag,
~~~S_3(x)=c^\dag(x)c(x)-1/2, ~~~n(x)=c^\dag(x)c(x) . \eqno(5) $$
It is convenient to replace the fermion $c(x)$ by $d(x)$, $c(x)=
e^{i\Gamma(x)}d(x)$, $c^\dag(x)=d^\dag(x)e^{-i\Gamma(x)}$.  In
addition, we choose $A(x,x+1)+\pi(n(x)-1)=\Gamma(x)-\Gamma(x+1)$.
As a result the Hamiltonian in Eq. (3) is replaced by:
$$\tilde{H}=-\frac{J_\perp}{2}\sum_{x=0}^{N-1}\left(d_+(x)d(x+1)+H.c.\right)
+J_\parallel\sum_{x=0}^{N-1}\left(n(x)-\frac{1}{2}\right)\left(n(x+1) 
-\frac{1}{2}\right)$$ 
$$+g\mu_Bh_0\sum_x\left(n(x)-\frac{1}{2}\right) . \eqno(6) $$ 
To obtain the form given in Eq. (6), we require that the phase 
$\Gamma(x)$ obeys the condition
$$e^{i[A(x,x+1)+\pi(n(x)-1)-\Gamma(x)+\Gamma(x+1)]}=1 . \eqno(7) $$

The periodicity of the spin operator $\vec{S}(x)=\vec{S}(x+L)$
together with the condition on the Berry connection gives rise to
a {\it twisted boundary condition} for the $d(x)$, $d^\dag(x)$
fermions:
$$d(x+N)=-e^{i\sum^{N-1}_{J=0}A(x+J,x+J+1)}e^{i2\pi N_F}
e^{-i\pi(N+1)}d(x) ,  \eqno(8a) $$ where
$N_F=\sum_{x=0}^{N-1}n(x)$ and $L=Na$.  Replacing the sum in Eq.
(8a) by a continuous integral we find, using Stokes theorem, that
$$\sum_{J=0}^{N-1}A(x+J,x+1+J)\simeq\oint A_\phi d\phi=\int_0^{\theta_0}
\int_{-\pi}^{\pi}\left(-\frac{\partial
A_\phi}{\partial\theta}\right) d\phi
d\theta=\pi(1-\cos\theta_0)\equiv\Omega(\theta_0) . \eqno(8b) $$
Here, $\Omega(\theta_0)$ measures the twist on the boundary
conditions.  Consequently, the equation for $d(x)$ is replaced by
$$d(x+L)=-e^{i\pi(\Omega(\theta_0)/\pi-1)}d(x) e^{i\pi N}=-e^{i\pi
(\delta-1)}d(x) , \eqno(8c) $$ where
$\delta=\Omega(\theta_0)/\pi+\Delta_N$, $\Delta_N=0$ if $N$ is
even and $\Delta_N=1$ if $N$ is odd.  Equation (8c) expresses the
twist of the boundary condition.  Consequently, the ``twist" of the
boundary condition is equal to the solid angle $\Omega(\theta_0)$, 
see Fig. 1.  Equations (8b) and (8c) represent one of our major 
results: a geometrical rotation of the spin-1/2 coordinate gives 
rise to a Berry phase which changes the boundary conditions for 
the wavefunction.

We construct the continuum representation of the Hamiltonian
$\tilde{H}$.  The fermion operator $d(x)$ is replaced by
$\psi(x)$:
$$ d(x)=\sqrt{a}\psi(x)=\sqrt{a}[e^{ik^0_F x}e^{-i(2\pi/L)(\delta-1)x}
R(x)+e^{-ik^0_F x}e^{-i(2\pi/L)(\delta-1)x}L(x) , \eqno(9) $$
where $k^0_F=\pi/2a$ and $R(x)$ and $L(x)$ are the right and left
moving operators at half filling:
$$R(x)=\frac{1}{\sqrt{2\pi a}}e^{i\alpha_R}e^{i(2\pi/L)(\hat{N}_R-1/2)x}
e^{i\sqrt{4\pi}\hat{\theta}_R} , $$
$$L(x)=\frac{1}{\sqrt{2\pi a}}e^{-i\alpha_L}e^{-i(2\pi/L)(\hat{N}_L-1/2)x}
e^{-i\sqrt{4\pi}\hat{\theta}_L} . \eqno(10)  $$

The representation given in Eq. (10) has been introduced in Refs.
[7,13]. $\hat{\theta}_R$ and $\hat{\theta}_L$ are the particle hole
excitations (non-zero modes), $\hat{N}_R$ and $\hat{N}_L$ are the
{\it zero mode} charge operators.  $\hat{N}_R$ and $\hat{N}_L$
measure the additional charge with respect to the half-filled
spinless Fermi sea.  $\alpha_R$ and $\alpha_L$ represent the zero
mode coordinates which are canonical conjugates to $\hat{N}_R$ and
$\hat{N}_L$, respectively:
$$[\alpha_R,\hat{N}_R]=[-\alpha_L,\hat{N}_L]=i, ~~~[\alpha,\hat{J}]=2i, ~~~
[\beta,\hat{Q}]=2i, ~~~\hat{J}=\hat{N}_L-\hat{N}_R,
~~~\hat{Q}=\hat{N}_L +\hat{N}_R, $$
$$\alpha=\alpha_R+\alpha_L, ~~~\beta=\alpha_L-\alpha_R . \eqno(11) $$

The complete set of zero mode eigenfunctions are given by
$$\langle\alpha_R|N_R\rangle=\frac{1}{\sqrt{2\pi}}e^{i\alpha_R N_R},
~~~ \langle\alpha_L|N_L\rangle=\frac{1}{\sqrt{2\pi}}e^{i\alpha_L
N_L} \eqno(12a) $$ or
$$\langle\alpha|J\rangle=\frac{1}{\sqrt{4\pi}}e^{i\alpha J/2}, ~~~
\langle\beta|Q\rangle=\frac{1}{\sqrt{4\pi}}e^{i\beta Q/2} ,
\eqno(12b) $$ where $\hat{N}_R|N_R\rangle=N_R|N_R\rangle$,
$\hat{N}_L|N_L\rangle=N_L N_L\rangle$,
$\hat{J}|J\rangle=J|J\rangle$ and $\hat{Q}|Q\rangle=Q |Q\rangle$.

We perform a unitary transformation which removes the chiral phase
$(2\pi/L)(\delta-1)x$ from Eq. (9).  The Fermi momentum
$k^0_F=\pi/2a$ is shifted to $k_F\ne k^0_F$ such that the vacuum
is shifted as a function of the magnetic field $h_0$.  We
introduce the new unitary transformation
$$ V=e^{-(i/2)[(\delta-1)-q]\alpha_R}e^{-(i/2)[(\delta-1)+q]\alpha_L} .
\eqno(13) $$ As a result of the unitary transformation $\psi(x)$
is replaced by
$$V^\dag\psi(x)V=e^{i(k^0_F+\pi q/L)x}R(x)+e^{-i(k^0_F+\pi q/L)x}L(x) .
\eqno(14) $$ The value of $q$ can be chosen such that the vacuum
is shifted and the term $g\mu_B h_0\sum(n(x)-1/2)$ is canceled
\cite{sun}.

The effect of the unitary transformation given in Eq. (13) is to
change the boundary condition for the eigenfunction in Eqs. (12),
$$V^\dag|J\rangle|Q\rangle=|J)|Q) , \eqno(15a) $$
such that $\langle\alpha+4\pi|J\rangle=\langle\alpha|J\rangle$ is
replaced by
$$\langle\alpha+4\pi|J)=e^{i(\delta-1)2\pi}\langle\alpha|J) .
\eqno(15b) $$ Similarly we have for $\langle\beta|Q)$,
$$\langle\beta+4\pi|Q)=e^{i(q-[q])2\pi}\langle\beta|Q) ,
\eqno(15c) $$ where $[q]$ is the closest integer to $q$.

Due to the boundary condition in Eq. (15b), the eigenvalues $J$
are given by $J=n/2+(\delta-1)/2$, $n=0,\pm1,\pm2,...$ and $Q=m/2
+(q-[q])/2$, $m=0,\pm1,\pm2,...$.  Using the unitary
transformation given in Eq. (13) and the Bosonic representation in
Eq. (10), we obtain the Bosonic form of the Hamiltonian:
$$\hat{H}=V^\dag\tilde{H}V=V^\dag\hat{U}^\dag H\hat{U}V, ~~~\hat{H}=
\hat{H}_{n=0}+\hat{H}_{n\ne0}+\Delta E_0 , \eqno(16a) $$ where
$\hat{H}_{n=0}$ is the zero mode Hamiltonian
$$\hat{H}_{n=0}=\frac{\pi v}{L}(\hat{N}^2_R+\hat{N}_L^2+2\gamma
\hat{N}_R\hat{N}_L) , \eqno(16b) $$ where $v\equiv J_\perp
a(1+J_\parallel/\pi J_\perp)$ and $\gamma= 2J_\parallel a/\pi v$.
$\hat{H}_{n\ne0}$ is the non-zero mode Bosonic Hamiltonian
\cite{sun}
$$\hat{H}_{n\ne0}=\int_0^L dx\left\{\frac{\hat{v}}{2}\left[K(\partial_x
\hat{\phi})^2+\frac{1}{K}(\partial_x\hat{\theta})^2\right]+\lambda\cos(2
\alpha+\sqrt{16\pi}\hat{\theta}+\frac{4\pi}{L}\hat{Q}x+4(K_F-K^0_F)x)
\right\} , \eqno(16c) $$ where $\hat{v}\equiv
v\sqrt{(1-\gamma^2)}$, $K=\sqrt{(1-\gamma)/(1+\gamma)}$,
$\lambda\equiv\hat{\lambda}/a^2$, $\hat{\lambda}\equiv
J_{\parallel}a/2 \pi^2$ with
$\hat{\phi}=\hat{\theta}_L-\hat{\theta}_R$, $\hat{\theta}=
\hat{\theta}_L+\hat{\theta}_R$.  Here $q$ is chosen such that the
linear term $g\mu_B h_0$ is canceled and the ground state energy
is shifted by,
$$\Delta E_0=-\frac{\pi L}{2a}\frac{1}{J_\perp}\frac{g\mu_B h_0}{1+
5J_\parallel/\pi J_\perp} , ~~~ 2\Delta K_F\equiv
2(K_F-K^0_F)\equiv2\pi q/L=-\frac{2\pi}{L}\frac{g\mu_B
h_0}{J_\perp(1+5J_\parallel/\pi J_\perp)} . \eqno(17) $$ Equation
(17) shows that the shift in the Fermi momentum depends on the
ratio between the Zeeman energy to the spin liquid lowest state
excitation energy.

The spin current is obtained from the Hamiltonian given in Eq.
(16).  The Heisenberg equation of motion for the zero modes
introduced in Ref. [13] allows us to construct the spin current
operator.  We replace the electric charge by the magnetic charge
$s=g\mu_B$.  As a result the spin current is given by the time
derivative of the zero mode coordinate:
$$I_M=\frac{s}{2\pi}\frac{d\alpha}{dt}=\frac{s}{i h}[\alpha,\hat{H}]
=\frac{sv_s}{L}(\hat{N}_L-\hat{N}_R), ~~~v_s\equiv\frac{J_\perp
a}{\hbar} \left(1-\frac{J_\parallel}{\pi J_\perp}\right).
\eqno(18) $$ From Eq. (18) we see that the spin current is
determined by the spin liquid velocity $v_s=J_\perp a/\hbar$ and
the expectation value of the current operator
$\hat{J}=\hat{N}_L-\hat{N}_R\equiv-i2d/d\alpha$.

To calculate the spin current we find the ground state
wavefunction given by the eigenstate of the zero mode Hamiltonian.
The zero mode Hamiltonian is obtained as a result of projecting
out the non-zero mode field $\hat{\theta}(x)$.  This projection is
obtained with the help of the Renormalization Group (R.G.)
equations given in Ref. [14].  We consider the mesoscopic region,
namely that the thermal length $L_T=\hbar v_s/k_BT$ is larger than
the circumference $L$ of the ring.  $L$ defines the ring temperature
$T^{ring}=\hbar v_s/k_BL$.  The backward coupling is renormalized
and replaced by $\lambda_{eff}$.  At a scale $b$ we have:
$\hat{\lambda}(b)\sim\hat{\lambda}b^{-x}$ where $1<b<L/a$ and
$x=2(2K-1)$. For $J_\perp> J_\parallel$, $x>0$ and $\lambda(b)$
decreases. For $J_\perp \sim J_\parallel$ we have $x\simeq0$ and
$\hat{\lambda}(b)\simeq \hat{\lambda}/(1+\hat{\lambda} ln(b))$
decreases much more slowly.

When $J_\parallel>J_\perp$, $\hat{\lambda}(b)$ grows with $b$ at
the scale $b^\ast\simeq e^{1/\hat{\lambda}(0)}$.  When $\hat{\lambda} 
(b)$ diverges, it signals that a {\it spin gap} is
opened.  Thus for a finite ring, {\it gapless excitations} are
possible in the regime $\pi J_\perp>J_\parallel>J_\perp$ if the
length of the ring $L$ is shorter than $\exp(1/\hat{\lambda})$,
$L<a\exp(1/\hat{\lambda})$, and the thermal length $L_T$ obeys
$L_T>\exp(1/\hat{\lambda})>L$.

Once the particle hole boson $\hat{\theta}(x)$ has been integrated
away, one obtains a zero mode Hamiltonian.  The zero mode
eigenfunctions $\psi_{n,m}(\alpha,\beta)$ of the Hamiltonian given
in Eq. (16) have the form:
$$\psi_{n,m}(\alpha,\beta)=Ae^{i(\delta-1)\alpha/2}e^{i(q-[q])\beta/2}
e^{in\alpha/2}e^{im\beta/2}\left[1+\sum_{r=\pm2,\pm4,...}C_r
e^{ir\alpha/2}\right] . \eqno(19) $$ For simplicity we consider the case
that $q$ is an integer, $q-[q]=0$.  Under this condition the
wavefunction in Eq. (19) is characterized by $n=0,\pm1,\pm2,...$
and $m=0$.

The effective zero mode Hamiltonian for $Q=0$ has the form:
$$\hat{H}_{eff}=\frac{\hbar \pi v_s}{2L}\hat{J}^2+\lambda_{eff}
\cos(2\alpha +2\pi q) , \eqno(20) $$ where
$\lambda_{eff}=(J_\parallel/2\pi)F(L/a)[\sin(2\pi q)/(2\pi q)]$.
Here $F(L/a)$ is the scaling function
$$F(L/a)=(L/a)^{-x},~~~x>0~~~and~~~F(L/a) =(1+\hat{\lambda}ln(L/a))^{-1},~~~
x\sim0 .$$ Here, $q$ measures the ratio of the Zeeman energy and
the spin liquid lowest energy state excitation in the ring.  When
$q\ll1$ ($h_0\rightarrow0$) the last term in Eq. (20) is replaced
by $\cos2\alpha$.  For large magnetic fields, $q\simeq0.5$ the
cosine term can be ignored and the spin current is given by
$I_M\simeq-g\mu_B (v_s/L)(1-\Omega(\theta_0)/\pi)$.  Using typical
values of exchange energy $J\sim100K-1000K$ and magnetic fields of
$5-10$ Tesla one finds $q\sim1$ justifying the neglect of the
Ising term.  In the remaining discussion we consider the
modification of the current caused by the term
$\lambda_{eff}\cos(2\alpha+2\pi q)$.

Using Bloch theory we compute the eigenfunctions of the
Hamiltonian in Eq. (20).  We find
$\psi_{n,l}(\alpha)=(1/\sqrt{4\pi})
\exp(i(\delta-1)\alpha/2)\exp(in\alpha/2)U_{n,l}(\alpha)$;
$n=0,1,2, 3$ are the Bloch states in the reduced zone and $l$ is
the energy band index.  We restrict the discussion to the sector
$Q=0$.  Therefore the possible states for ${J}$ are $n=0,2$.  Odd
states are excluded since they give states with $Q\ne0$.  Using
Bloch states $\psi_{n,l} (\alpha)\equiv\langle\alpha|n,l)$ with
eigenvalues $E_{n,l}$, we compute the spin current formula using
Eq. (18) and find
$$I_M=-g\mu_B\frac{v_s}{L}\sum_{n=0}^2\sum_l\frac{1}{Z}(n,l|\hat{J}|
n,l)e^{-E_{n,l}/k_BT}=-2g\mu_B\sum_{n=0}^2\sum_l\frac{1}{Z}\left(\frac
{\partial E_{n,l}}{\partial n}\right) e^{-E_{n,l}/k_BT} .
\eqno(21) $$ Here $Z$ is the zero mode partition function
$Z=\sum_{n=0}^2\sum_l e^{-E_{n,l}/k_BT}$ and $|n,l)$ are Bloch
states with the eigenvalues $E_{n,l}$.  The second part of Eq.
(21) was obtained from Bloch's theory, see Ref. [15].  Using Eq.
(21) we calculate the current for two cases: (a) weak coupling,
$\lambda_{eff}/(\hbar\pi v_s/2L)\ll1$, and (b) strong coupling
$\lambda_{eff}/(\hbar\pi v_s/2L)\gg1$.

{\it (a) The weak coupling case:} Because $\lambda_{eff}$
decreases as $L^{-x}$, $x<1$ and the kinetic term (the term
proportional to $\hat{J}^2$) decreases as $1/L$, the weak coupling
region is achieved only for $J_\perp\gg J_\parallel$ or large
magnetic fields $|\sin2\pi q/2\pi q|\ll1$.  For this case we can
use first order perturbation theory [16] to find the ground state
Bloch wavefunction $\psi_{n=0}(\alpha)$ for the Hamiltonian in Eq.
(20):
$$\psi_{n=0}(\alpha)=e^{i(\delta-1)\alpha/2}\left[1+\frac{\lambda_{eff}L}
{\pi v_s}\left(\frac{e^{-i2\alpha}e^{i2\pi q}}{2-(\delta-1)}-\frac
{e^{i2\alpha}e^{-i2\pi q}}{(\delta-1)+2}\right)\right] . \eqno(22)
$$ Substituting Eq. (22) in the first part of Eq. (21) gives for
the spin current
$$I_M=-g\mu_B\frac{v_s}{L}\left(1-\frac{\Omega(\theta_0)}{\pi}\right)
\left[1-\frac{1}{(4\pi^3)^2}\left(\frac{J_\parallel/J_\perp}{1-
J_\parallel/\pi J_\perp}\right)^2\left(\frac{L}{a}\right)^{2(1-x)}
\left(\frac{\sin2\pi q}{2\pi q} \right)^2 \right] , \eqno(23) $$
where $v_s=2J_\perp a(1-J_\parallel/\pi J_\perp)$ and $1>x>0$.
Equation (23) is valid for values of $J_\parallel/J_\perp$
satisfying $(J_\parallel/J_\perp)|\sin2\pi q/2\pi q|<(a/L)^{1-x}$.

{\it (b) The strong coupling case:} For this case we compute the
eigenfunctions of Eq. (20) using the ``instanton" method given in
Ref. [17] or the solution given on page 231 in Ref. [15].  For
large $\lambda_{eff}$ the eigenvalues $E_{n,l=\pm}$ are given by
$E_{n,\pm}=(\hbar\omega/2)\pm(\hbar\omega \sqrt{D}/\pi)\cos[2\pi
n+2\pi(\delta-1)]$, where the energy scale $\hbar\omega$ is
determined by $\hbar\omega\equiv(\hbar v_s^{(0)}/L)
[(1/2\pi)(J_\parallel/J_\perp)(L/a)^{1-x}|\sin2\pi q/2\pi q|]$
with $v_s^{(0)} =J_\perp a/\hbar$ and the band width $\sqrt{D}$ is
given by $\sqrt{D}
\equiv[\exp{-{\omega}\int_0^{2\pi}\sqrt{2\cos(\alpha +2\pi q)}
d\alpha}]$.  Using the spin current given by Eq. (21) we find,
$$I_M=-g\mu_B\frac{v_s^{(0)}}{L}\sin[2\pi(1-\Omega(\theta_0)/\pi]
\sqrt{D}\sqrt{(8/\pi)(J_\parallel/J_\perp)|sin(2\pi q)/2\pi q|
(L/a)^{1-x}} . \eqno(24) $$ From Eq. (24) we see that when
$J_\parallel/J_\perp$ increases the current is suppressed since
the band width $\sqrt{D}$ scales like
$\exp(-\sqrt{J_\parallel/J_\perp})$.  The results in Eq. (23) and
(24) are a consequence of the twist of the boundary condition
obtained in Eq. (8c).  In both cases the current was obtained for
the even case, namely $\Delta_N=0$.  For the odd case $\Delta_N=1$
and instead of $1-\Omega(\theta_0)/\pi$ we have
$\Omega(\theta_0)/\pi$.

Whether it is possible to detect the spin liquid current is an
important question.  Because of the Aharanov-Casher effect, when a
metallic rod is inserted at the center of the ring one expects to
find an electrostatic potential $\Phi_E(0)\simeq(I_M^{(0)}/
L)(2\pi)^2$, i.e., a voltage proportional to the spin current
$I_M^{(0)}$.  For a magnetic field $h_0\simeq10$ Tesla, an
exchange energy $J_\perp\sim2000$ Kelvin and a ring of 10 $\mu$m,
the Zeeman/exchange energy ratio is of the order of 10 allowing us to
neglect the Ising term $J_\parallel$ and obtain a maximal current
$I_M^{(0)}$.  For a ring of 10 $\mu$m we need to have a low
temperature such that $L_T= \hbar v_s/k_BT>10$ $\mu$m;
corresponding to a temperature of one Kelvin.

Materials that can be used to fabricate antiferromagnetic spin-1/2
ring structures include copper-benzoate.  It is possible to
fabricate such rings using plasma etching techniques in which a
mesoscopic ring structure is etched out on a substrate and the
desired magnetic material is deposited creating a magnetic ring
structure.  The persistent spin current can be measured using a
scanning capacitance probe.  Due to spin-orbit coupling the
current will generate an electric field and therefore a potential
difference.  A small current in the ring will generate a voltage
of the order of a nano-volt.  An alternative approach is to attach
two leads to a Luttinger liquid ring in the Aharonov-Bohm geometry
with a crown shaped magnetic field.  This generates a phase both
in spin and charge.  If the ring in the center is threaded with a
flux tube \cite{kampf} that exactly cancels the magnetic flux due
to the crown shaped field, the remaining contribution is due to
spin. Note that this flux tube will not affect the magnetic field
on the ring.

In conclusion, we have shown that a crown shaped magnetic field
when acting on a spin-1/2 Heisenberg antiferromagnetic chain can
give rise to a spin current proportional to the solid angle
$\Omega(\theta_0)=\pi(1- \cos\theta_0)$.  The amplitude of the
current is controlled by the ratio $J_\parallel/J_\perp$ and the
Zeeman energy.  A possible experimental realization has been
suggested.

{\bf Acknowledgement:} D.S. was supported by the U.S. Department
of Energy grant number FG02-01ER45909.  This work was supported by
the Los Alamos Laboratory Directed Research and Development
Program.

\begin{figure}
\caption{The geometry of a spin ring with circumference $L$ in a crown 
shaped magnetic field $\vec{h}(x)$.  The coordinate $x$ is along the 
ring.  The field has a tilt angle of $\theta_0$ and subtends a solid 
angle $\Omega(\theta_0)$. }
\end{figure}

\end{document}